\documentclass[pre,twoside,twocolumn,showpacs,byrevtex,superscriptaddress]{revtex4}

\newcommand{\Prn}{P(r; n)}
\newcommand{\Nrn}{N(r; n)}
\newcommand{\Pins}{P_{\rm ins}(r; n)}

\newcommand{\alphamf}{\alpha_{\rm mf}}

\newcommand{\growthratei}{G_i}
\newcommand{\injrate}{\kappa}

\newcommand{\lambdamax}{\lambda_{\rm max}}

\newcommand{\porejamtime}{\tau_{\rm jam}}

\def\d{{\rm d}}

\usepackage{graphicx,epsfig,verbatim,enumerate}
\usepackage{amssymb}
\usepackage{ifthen}
\newboolean{twocolswitch}

\newcommand{\etal}{\textit{et al.}}
\newcommand{\www}[1]{\url{#1}}
\newcommand{\req}[1]{(\ref{#1})}

\newcommand{\dee}[1]{\mbox{d}#1}

\newcommand{\avg}[1]{\left\langle#1\right\rangle}

\setboolean{twocolswitch}{true}

\begin{document}

\title{
Packing-Limited Growth
}

\markboth{PACKING LIMITED GROWTH}{PETER SHERIDAN DODDS AND JOSHUA S.\ WEITZ}

\author{
  \firstname{Peter Sheridan}
  \surname{Dodds}
  }
\email{dodds@ldeo.columbia.edu}
\affiliation{
        Columbia Earth Institute, 
        Columbia University,
        New York, NY 10027.
        }

\author{
  \firstname{Joshua S.}
  \surname{Weitz}
  }
\email{jsweitz@segovia.mit.edu}
\thanks{(Please direct correspondence to both authors.)}
\affiliation{
        Department of Earth, 
        Atmospheric and Planetary Sciences,
        Massachusetts Institute of Technology, 
        Cambridge, MA 02139.
        }
\affiliation{
        Department of Physics,
        Massachusetts Institute of Technology,
        Cambridge, MA 02139.
        }

\date{\today}

\begin{abstract}
We consider growing spheres seeded by 
random injection in time and space.
Growth stops when two spheres meet leading eventually to a jammed state.
We study the statistics of growth limited by packing
theoretically in $d$ dimensions and
via simulation in $d=2$, $3$, and $4$.
We show how a broad class of such models 
exhibit distributions of sphere radii with
a universal exponent.  
We construct a scaling theory that relates the fractal
structure of these models to the decay
of their pore space, a theory that
we confirm via numerical simulations.  
The scaling theory also predicts an upper 
bound for the universal exponent and is 
in exact agreement with numerical results
for $d=4$.
\end{abstract}


\maketitle

\section{Introduction}
\label{manna.introduction}

The dynamic packing of objects is 
an often overlooked variation on the theme of static packing.
Given a mechanism for the creation, growth, movement, and
interaction of like objects in a given dimension, 
what structures result?
Here, we find universal features of a simple yet
broad class of such mechanisms
falling under the rubric ``packing-limited growth'' (PLG).
A PLG mechanism entails objects being seeded randomly, 
growing according to a rule that may
be specific to each object, and stopping when a part of their boundary
hits that of another object.  
A motivating physical example of this kind of pattern formation 
may be found in the competition between tree crowns
in dense forests~\cite{horn1971, takahashi1996}, the structure of
porous media~\cite{vanderMarck1995,turcotte1997,hecht2000}, 
and the generalized problem of dense packings~\cite{conway1999}.

A simple model of PLG
has previously been studied by 
Andrienko, Brilliantov, and Krapivsky~\cite{andrienko94,brilliantov94}
(the ABK model).
In their setting, spheres are seeded randomly in space
and time.  A sphere's radius increases linearly with time, 
halting when another sphere is touched.  
This model is amenable
to an approximate analysis for $d>1$~\cite{brilliantov94}
and has an exact solution in $d=1$.
In this present work, we are interested
in the limiting distribution of radii, $N(r)$.
For PLG models, we expect
$N(r) \propto r^{-\alpha}$ for small $r$.
Note that the fractal dimension $D$ of the set
comprising all sphere centers, which is often
measured instead of $\alpha$, is related to $\alpha$
as $D=\alpha-1$~\cite{manna92a}.
In $d=1$, the exact solution
gives $\alpha=1$, which corresponds
to $D=0$ (meaning the number of centers
diverges logarithmically)~\cite{brilliantov94}.
For $d=2$, Andrienko \etal~\cite{andrienko94}
report that $D \simeq 1.75$
and hence $\alpha \simeq 2.75$ based on numerical evidence.
Elsewhere, the same authors~\cite{brilliantov94} 
determine numerically that 
$\alpha \simeq 2.53$.

We claim that the actual value of $\alpha$
is essentially independent of the specifics of the growth dynamics.
To see why, we examine a related $d=2$ 
random packing model due to Manna~\cite{manna92a}.
The packing process begins in a finite-sized volume 
(or alternatively with an initial population of fixed radius disks).
New disks are added one at a time by randomly choosing a
point in the packing's pore space and centering there the largest possible
non-overlapping disk.  Manna finds $\alpha \simeq 2.62$
and $\alpha \simeq 2.64$ for two example packings.

We refer to the Manna model as ``random Apollonian packing'' (RAP)
since it may be seen as a variation of 
the well known Apollonian packing~\cite{mandelbrot83}.
For the $d=2$ version of the latter, 
pore spaces are always formed by three disks each touching
the other two and disks are added so as to fill the pore space
fully (i.e., the inserted disk touches all three of its surrounding
neighbors).  
Numerics give
$\alpha \simeq 2.31$ 
for Apollonian packing in agreement with analytically 
derived bounds~\cite{manna91}.
Aste~\cite{aste96} has shown for static polydisperse
packings that are space filling,
Apollonian packing provides the lowest
bound on $\alpha$ while the upper bound is $\alpha=d+1$.

There are two important observations to make here.
First, as noted by Brilliantov~\etal~\cite{brilliantov94},
the RAP model is the ABK model in
the case of infinitely fast growth:
as soon as a sphere
is nucleated, it instantaneously expands to
hit the nearest sphere boundary.
Second, for general PLG
models, pore spaces evidently 
increase in number and decrease in size with time.
This means that the likelihood of two spheres nucleating
in the same pore space also decreases.
Eventually the mechanism 
of the RAP model must take over,
and all collisions will be between a growing sphere in a pore
space and a ``stuck'' sphere forming part of the pore
space boundary.  
Thus, the RAP model is not just
a curious end point of the ABK model but, in fact,
entails the sole mechanism describing how small
radius spheres pack.  
We suggest then that
measurements of $\alpha$ in the ABK model
should coincide with revised estimates from simulations
of the RAP model.  
In the remainder of the paper, we describe a general class of 
PLG models for which 
$\alpha$ is invariant, 
provide a simple solution for the $d=1$ problem,
develop a scaling theory that 
describes how the radii distribution and volume of pore space evolve with time,
and provide results from extensive numerical simulations.

\section{General model of packing-limited growth}
\label{manna.general}

\begin{figure}[tbp]
  \begin{center}
    \epsfig{file=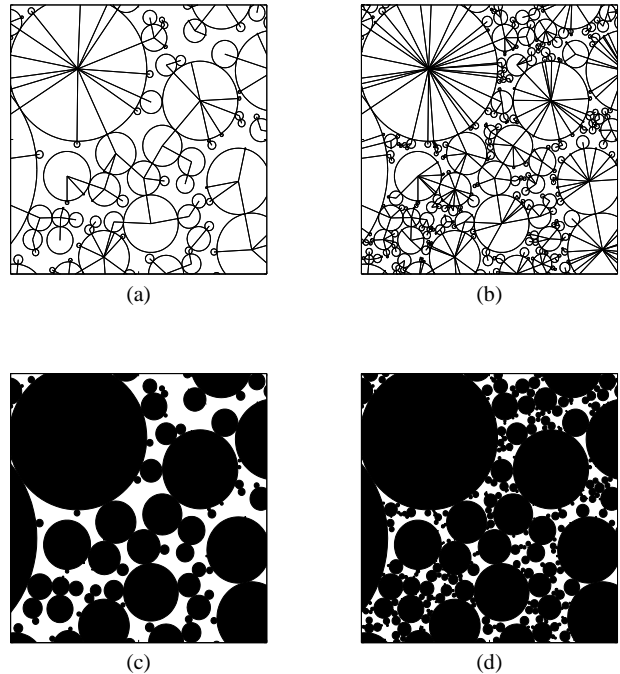,width=0.95\columnwidth}
    \caption{
      The dynamics and limiting states of PLG models.  In (a) and (b),
      collisions between centers are marked with lines while
      in (c) and (d),
      the region contained by spheres is black and
      the pore space is left white.
      All four pictures come from
      the same simulation, (a) and (c) depict the condition of a subsection
      of a system packed with 500 spheres, (b) and (d) depict the
      same subsection after 2000 spheres have been packed.
      }
    \label{fig:manna.spindle}
  \end{center}
\end{figure}

For our general conception of PLG
we take $d$-dimensional spheres growing
in a volume $V$.  
Spheres are nucleated
randomly in space at a rate $\injrate$ per 
unit volume surviving only if they are injected
into the unoccupied pore space.
All spheres stop growth upon contact with a neighboring sphere.
Figure~\ref{fig:manna.spindle} provides a visual context
for the dynamics explained below.
The $i$th sphere, which is initiated at $t_{i}$, grows at 
a rate $\growthratei(t-t_{i})$ giving the radius $r_i(t)$ as
\begin{equation}
  \label{eq:manna.growthrate}
  r_i(t) = \int_{0}^{t-t_{i}} \growthratei(u)\, \dee{u},
\end{equation}
for $t \ge t_{i}$. 
We assume the weak requirement 
that each sphere grows in a strict monotonic fashion,
i.e., for each $i$, $\growthratei(t-t_i) \geq \epsilon>0$.
The model is thus one of 
arbitrary individual growth limited solely by packing.
The dynamics continue in the above fashion until the volume $V$
is completely filled and a final jammed state is reached.

Consider an individual pore $\Gamma$ of diameter $\lambda$ (i.e., the maximum
separation of two boundary points that can be joined
by a line contained entirely within the pore).
The rate $\nu$ at which
spheres nucleate in $\Gamma$ (excluding growth for the moment)
is given by 
\begin{equation}
  \label{eq:manna.kappap}
  \nu \simeq \kappa V_d \lambda^d,
\end{equation}
where $V_d$ is the volume of $d$-dimensional
sphere of unit radius.  The typical
time $\tau$ between nucleations in $\Gamma$ is therefore
\begin{equation}
  \label{eq:manna.tp}
  \tau = 1/\nu \simeq \kappa^{-1} V_d^{-1} \lambda^{-d}.
\end{equation}
On the other hand, if sphere $i$ nucleates in $\Gamma$,
it will jam in a time $\porejamtime$ 
that can be 
no greater than the time it takes for its radius to reach $\lambda/2$.
Together with the assumption
that $\growthratei(t-t_i) \geq \epsilon>0$, this gives
\begin{equation}
  \label{eq:manna.tj}
  \porejamtime \lesssim \lambda/2\epsilon.
\end{equation}
So when $\porejamtime \ll \tau$, i.e., when
\begin{equation}
  \label{eq:manna.tj2}
  V_d \lambda^{d+1} \kappa/2\epsilon \ll 1,
\end{equation}
we expect the packing mechanism
to be the same 
as the RAP model---all spheres 
will be stopped by an existing stopped sphere
and never by another moving sphere.
Growth rate therefore
becomes irrelevant as far as the final packing is concerned.
From Eq.~\req{eq:manna.tj2},
we see that the general model reduces to the RAP model
when the maximum pore size $\lambdamax$ satisfies
\begin{equation}
  \label{eq:manna.lambdamax}
  \lambdamax \ll
  \left(2\epsilon/V_d\kappa\right)^{1/(d+1)}.
\end{equation}
Since $\lambdamax$ steadily
decreases with $t$ (i.e., space fills up)
this condition will always be eventually satisfied.

\section{Exact $d=1$ solution}
\label{manna.d=1}

As we have noted, the ABK model has been solved exactly in
the $d=1$ case~\cite{brilliantov94} with the outcome
being $\alpha=1$.  
Here, we achieve the same result with a 
considerably simpler calculation.
Since we have posited that 
$\alpha$ is a universal exponent for a general class of 
packing-limited growth models, we may
choose the straightforward example of the $d=1$ RAP model.
We take the unit interval $[0,1]$ as the initial vacant pore space.
Spheres are now line segments and are limited either on
the left or right as they are added with the points 0 and 1 providing initial 
boundaries.  Thus, the unit interval is filled in from each end
with the one solitary pore space diminishing in the middle.
Writing the length of the $n$th line segment as $l_n$ we have
\begin{equation}
  \label{eq:manna.ln+1}
  l_{n} = z_{n-1}
  \left(
    1-\sum_{i=1}^{n-1} l_i
  \right),
\end{equation}
where $z_i$ is a random number uniformly distributed on the unit interval
and $l_1 = z_0$.  In other words, each new line segment is 
a random fraction of the current pore space.  
Iteratively solving
Eq.~\req{eq:manna.ln+1} gives
\begin{equation}
  \label{eq:manna.lnsol}
  l_{n} = z_{n-1}
  \prod_{i=0}^{n-2} (1-z_{i}).
\end{equation}
We can in principle find the distributions of the $l_i$
but for our present purposes their means are
sufficient and we have
\begin{equation}
  \label{eq:manna.lnmean}
  \avg{l_{n}} = 2^{-n}.
\end{equation}
We therefore expect the typical number of line segments
with $l' \geq l = 2^{-n}$ to be
\begin{equation}
  \label{eq:manna.lcdf}
  N(l'\geq l = 2^{-n}) = n = -\frac{\ln{l}}{\ln{2}}.
\end{equation}
Since this is the cumulative frequency distribution, 
i.e., $N(l' \geq l ) = \int_{l}^{\infty} N(l')\dee{l'}$,
we find the frequency distribution $N(l)$ must behave as
\begin{equation}
  \label{eq:manna.Nl1d}
  N(l) \simeq \frac{1}{\ln{2}} l^{-1},
\end{equation}
yielding, as expected, $\alpha=1$.

\section{Description of insertion probability and pore space volume}
\label{manna.Pinsert}

\begin{figure}[tbp]
  \begin{center}
    \epsfig{file=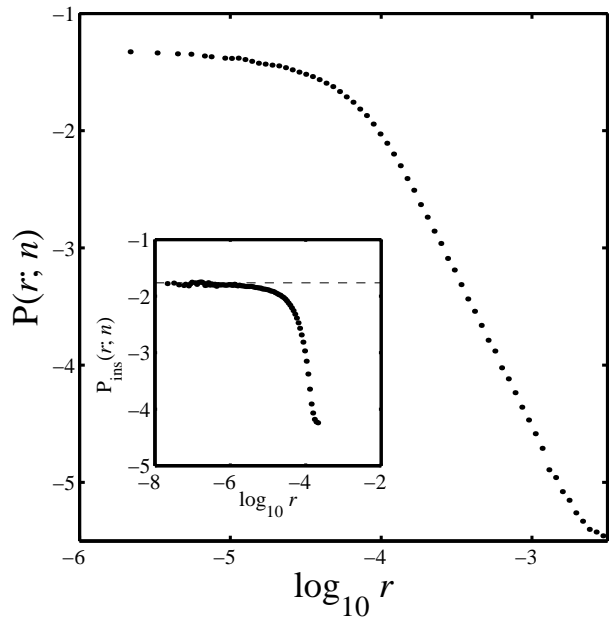,width=0.95\columnwidth}
    \caption{
      $P(r;n)$ for a $d=2$ RAP simulation with $n=10^6$.  
      The inset distribution
      is $\Pins$ which is obtained by simulating the RAP procedure
      for another $10^6$ disks without actually inserting any of them into the 
      pore space.  The dashed line in the inset figure corresponds to
      the theoretical prediction $P_{\rm ins}(0;n)=S(n)/\Phi(n)$.
      }
    \label{fig:manna.pinsert}
  \end{center}
\end{figure}

We next investigate the form of $\Pins$,
the probability distribution 
of the $(n+1)$th sphere's radius to 
be inserted into a packing.  
In addition, we characterize
$P(r; n)$, the probability distribution of
sphere radii after $n$ spheres have been packed.
Note that since we have shown
that a general class of PLG models reduce to the RAP model,
we now use the number of
spheres $n$ rather than time $t$ to
index our quantities.

We first note that the distribution $\Prn$ fills
in from the right with increasing $n$.  As the packing
fills in space, the maximum pore size either decreases 
or does not change and so does the 
maximum size of any sphere that may be added.
We write the radius of this largest sphere as $r_c$.
We observe that to a first approximation,
$\Prn$ above this cutoff scale $r_c$
follows its limiting power law form
while below the distribution it is essentially flat 
(see Fig.~\ref{fig:manna.pinsert}).
For the purposes of estimation,
we assume this form exactly as
\begin{equation}
  \label{eq:manna.Prn}
  \Prn = 
  \left\{
    \begin{array}{ll}
      \frac{\alpha-1}{\alpha} r_c^{-1} & \mbox{for $r<r_c$}\\
      \frac{\alpha-1}{\alpha} r_c^{-1} 
      \left(\frac{r}{r_c} \right)^{-\alpha} & \mbox{for $r \geq r_c$}.
    \end{array}
  \right.
\end{equation}
The corresponding frequency distribution is $\Nrn \propto \Prn$.
Since the tail of $\Nrn$ remains fixed as $n$ increases,
we can obtain the scaling of $r_c$ with $n$.
From Eq.~\req{eq:manna.Prn},
the tail of $\Nrn$ behaves as 
\begin{equation}
  \label{eq:manna.Ntail}
    \Nrn =  \frac{\alpha-1}{\alpha} n r_c^{\alpha-1} r^{-\alpha} = k r^{-\alpha},
\end{equation}
for $r > r_c$.  Since the prefactor $k$ must be constant,
we have 
\begin{equation}
  \label{eq:manna.rcn}
  r_c = 
  \left(
    \frac{\alpha-1}{\alpha k}
    n 
  \right)^{-1/(\alpha-1)}
  \propto n^{-1/(\alpha-1)}.
\end{equation}

The uniformity of $\Prn$ for $r<r_c$ closely
relates to the form of $\Pins$.  It is possible
to write down an exact expression for $\Pins$, that is,
\begin{equation}
  \label{eq:manna.Pins_exact}
  \Pins = \frac{\int \d V \delta(D_n(\vec{x})-r)}{\int \d V},
\end{equation}
where the integrals are over the pore space, and $D_n(\vec{x})$
is the distance from the point $\vec{x}$ to the closest pore
space boundary after $n$ spheres have been inserted.  This integral
may be solved exactly in the limit of very small radii,
\begin{equation}
  \label{eq:manna.Pins_exact2}
  \lim_{r\rightarrow 0} \Pins  = \frac{S(n)}{\Phi(n)},
\end{equation}
where $S(n)$ and $\Phi(n)$ are the surface area and
available pore space of the existing $n$ packed spheres.
This means that the region of pore space available for
insertion of an infinitesimal sphere 
is proportional to the surface area of the extant packing.
Assuming that this holds approximately for all spheres below the cutoff,
the full distribution $\Pins$  may be modeled
as a purely flat distribution,
\begin{equation}
  \label{eq:manna.Pins}
  \Pins = 
  \left\{
    \begin{array}{ll}
      r_c^{-1} & \mbox{for $r<r_c$}\\
      0 & \mbox{for $r \geq r_c$}.
    \end{array}
  \right.
\end{equation}
Good agreement with this approximation
is shown in the inset of Fig.~\ref{fig:manna.pinsert}.
Comparing Eq.~\req{eq:manna.Pins_exact2} 
with Eq.~\req{eq:manna.Pins},
we see that $r_c = \Phi(n)/S(n)$.
We calculate $S(n)$ and $\Phi(n)$ using $\Nrn$,
\begin{equation}
  \label{eq:manna.S(n)}
  S(n) = \frac{k\alpha V_d}{\alpha-d} r_c^{-(\alpha-d)},
\end{equation}
and
\begin{equation}
  \label{eq:manna.Phi(n)}
  \Phi(n) = \frac{k \alpha V_d}{(d+1)(d+1-\alpha)} r_c^{d+1-\alpha},
\end{equation}
where as before $V_d$ is the volume of a unit radius sphere
in $d$ dimensions.  Note that $S(n) \rightarrow \infty$ 
and $\Phi(n) \rightarrow 0$ as $n \rightarrow \infty$.

Using Eqs.~\req{eq:manna.S(n)} and~\req{eq:manna.Phi(n)},
and the result $r_c = \Phi(n)/S(n)$, we find an estimate of
$\alpha$ as
\begin{equation}
  \label{eq:manna.alphaest}
  \hat{\alpha} = d + \frac{d+1}{d+2}.
\end{equation}
In contrast,
a modified calculation
for the ABK model finds~\cite{brilliantov94}
\begin{equation}
  \label{eq:manna.alphamf}
  \alphamf = 1 + 
  d \left( 1 + 
    \exp\left[2 - \frac{2^{d+2}-2}{d+2}\right]\right).
\end{equation}
Note that $\hat{\alpha}$ 
must be an upper bound on the true value of $\alpha$
for normal Apollonian packing.
Furthermore, due to the nature of the approximations made
in forming $\Pins$ and $\Prn$, 
$\hat{\alpha}$ is also an upper bound for the exponent of the RAP model.
As we show in the following section when we consider our numerical 
results, both of these predictions of $\alpha$ appear to
hold for certain (mutually exclusive) ranges of $d$.

Using Eqs.~\req{eq:manna.rcn}, \req{eq:manna.S(n)},
and~\req{eq:manna.Phi(n)}, we are able to find the scalings of surface 
area and pore space volume with $n$,
\begin{equation}
  \label{eq:manna.S(n)scaling}
  S(n) \propto n^{-\gamma} \propto n^{(\alpha-d)/(\alpha-1)},
\end{equation}
and
\begin{equation}
  \label{eq:manna.Phi(n)scaling}
  \Phi(n) \propto n^{-\beta} \propto n^{-(d+1-\alpha)/(\alpha-1)}.
\end{equation}
These relations provide us with further
methods for determining $\alpha$ and for testing
scaling relations between the iterative
and structural nature of the packing.
Equation~\ref{eq:manna.Phi(n)scaling}, in particular,
affords a more robust measurement than
does obtaining $\alpha$ directly from $\Prn$,
and we employ this fact in our numerical investigations.

Note that for comparison between the current theory and that 
of ABK we must convert scaling predictions that depend on time to those that
depend on sphere number.
Specifically,
the ABK theory predicts $\Phi(t)\propto t^{-A}$
where $A=\exp[2-(2^{d+2}-2)(d+2)]$. 
Because the ABK
model can be seen as an RAP model with
infinite growth velocity, the time between events is inversely
proportional to the pore space available (given a uniform
rate for attempted nucleation $\kappa$).  So the result
for the decay of pore space from ABK may be rewritten 
as $\Phi_{ABK}(n)\propto n^{-\beta'}$, where
\begin{equation}
t_n=\frac{1}{\kappa}\sum_{i=1}^n\frac{1}{\Phi_{ABK}(i)}.
\end{equation}
This implies that $t_n\propto n^{1+\beta'}$.  So if
ABK predicts $\Phi(t)\propto t^{-A}$, this is equivalent
to writing $\Phi_{ABK}(n)\propto n^{-A(1+\beta')}$.  Equating
the two forms leads to the conclusion that 
\begin{equation}
\beta'=\frac{A}{1-A}.
\label{eq:manna.abktheory}
\end{equation}
This will be useful in the following section when the
scaling of pore space decay is examined.
Finally, key scaling relations are summarized
in Table~\ref{tab:manna.scaling}.

\begin{table}[htbp!]
 \begin{center}
   \begin{tabular*}{\columnwidth}{@{\qquad}l@{\extracolsep{\fill}}c@{\qquad}}
     \toprule
     relation & exponent \\ \hline 
     $P(r) \propto r^{-\alpha}$ &  $\alpha$ \\
     $r_c(n) \propto n^{-\delta}$  & $\delta = d+1-\alpha$ \\
     $S(n) \propto n^{\gamma}$ &  $\gamma = (\alpha-d)/(\alpha-1)$ \\
     $\Phi(n) \propto n^{-\beta}$ & $\beta = (d+1-\alpha)/(\alpha-1)$ \\
     \botrule
   \end{tabular*}
   \caption{
     Scaling relations between exponents for various parameters
     valid for general packing-limited growth models:
     $P(r)$ is the probability distribution of radii,
     $r_c$ is the radius of the typical largest sphere that
     may be inserted, $S(n)$ is the surface area of $n$ packed spheres,
     and $\Phi(n)$ is the pore space volume.
     All exponents are given in terms of $\alpha$ and the 
     dimension $d$.
     }
   \label{tab:manna.scaling} 
 \end{center}
\end{table}

\section{Numerical results}
\label{manna.numerics}

The numerical procedure for all variants of PLG
schemes obey the same basic algorithm.  Spheres are seeded
according to a Poisson distribution, with rate $\injrate$
and only grow within the pore space.  Once injected,
spheres grow according to model rules 
(linear velocity, heterogeneous, exponential, etc.)
until they collide and are jammed.  In the case of the 
RAP model, only one sphere is allowed to nucleate
and expand at any given point
in the simulation.  Using this event driven procedure we
are able to calculate limiting states and determine improved
estimates of the universal value of $\alpha$ 
for $d=2$, $3$, and $4$.  All packings are simulated on
a unit hypercube with periodic boundary conditions.
\begin{figure}[tbp]
  \begin{center}
    \ifthenelse{\boolean{twocolswitch}}
    { 
      \epsfig{file=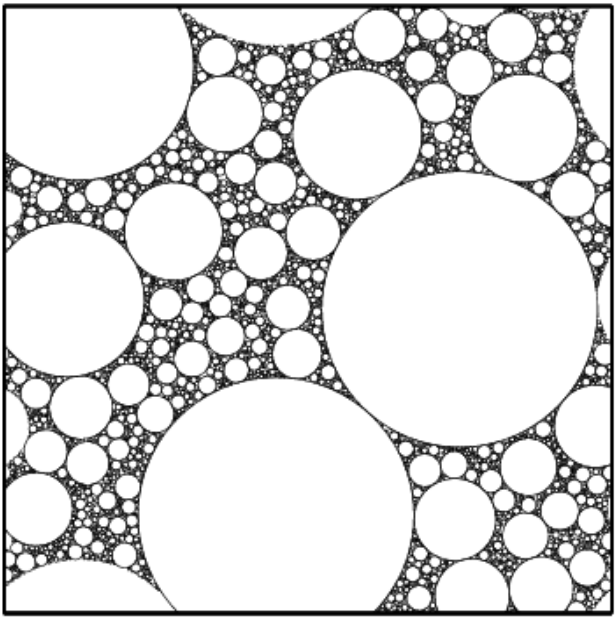,width=0.95\columnwidth}
      }
    { 
      \epsfig{file=fig3_basiccircles.ps,width=0.95\columnwidth}
      }
    \caption{
      A random Apollonian packing initially 
      seeded with two circles after 10,000 circles
      have been placed.  
      The density of the system is $\rho\simeq 0.94$.
      }
    \label{fig:manna.2dpic}
  \end{center}
\end{figure}
Figure~\ref{fig:manna.2dpic} shows an example packing using
the RAP model in $d=2$.  It is clear that the number of 
spheres will increase without bound and that the pore space
ultimately vanishes.  
The associations with the Apollonian case are visually evident.

\begin{table}[hbtp!]
 \begin{center}
   \begin{tabular*}{\columnwidth}{c@{\extracolsep{\fill}}cccc@{\quad}}
     \toprule
     & RAP & Heterogeneous & Exponential & Linear \\
     \hline
     $\injrate$ & n/a & 10 & 10 & $10^{-5}$ \\
     $G_i(t-t_i)$ & $\infty$ & 0.2 -- 2.0 & $e^{t-t_i}$ & 0.2 \\
     $N_{\mbox{s}}$  & 50    & 48               & 10                 & 1 \\
     $N_{\mbox{tot}}$ & $2.5\times 10^6$ & $2.3\times 10^6$ & $5.4\times 10^5$ & $3.4\times 10^5$ \\
     $\rho$ & 0.97 & 0.96 & 0.91 & 0.97 \\
     \botrule
   \end{tabular*}
   \caption{
     Simulation details for four different PLG models.
     $\kappa$ is the per unit volume sphere nucleation rate,
     $G_i(t-t_i)$ is the growth rate of the the $i$th sphere
     which is nucleated at time $t_i$, 
     $N_{\mbox{s}}$ is the number of simulations, 
     $N_{\mbox{tot}}$ is the total number of spheres placed, 
     and $\rho$
     is the approximate limiting density of each simulation.
     The sphere radii distributions for these models
     are shown in Figure~\ref{fig:manna.multilogcollapse3}.
     }
   \label{tab:manna.simdetails}
  \end{center}
\end{table}

\begin{figure}[tbp]
  \begin{center}
    \epsfig{file=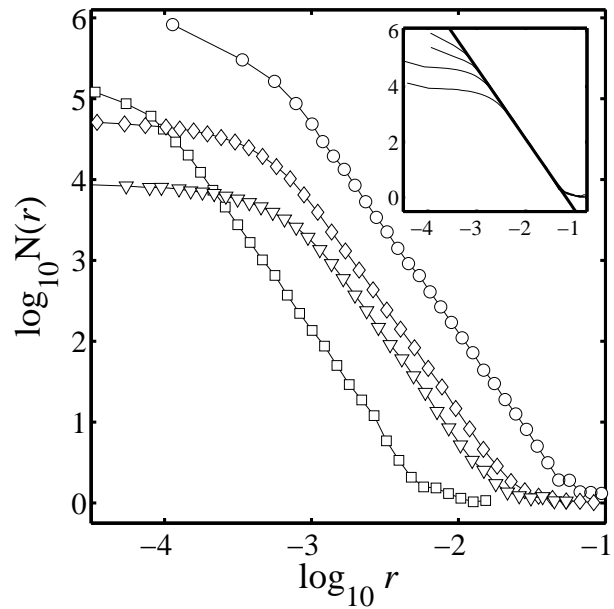,width=0.95\columnwidth}
    \caption{
      The form of $N(r)$ for four models: random Apollonian packing (circles),
      heterogeneous linear growth rates (diamonds), exponential growth
      rates (triangles), and linear growth rates (squares).
      Simulation details are contained in the text.
      The inset double-logarithmic
      plot shows $N(r)$ versus $r$ for each model
      shifted horizontally for the purpose of illustration.
      The slope of the solid straight line indicates the $d=2$ universal
      exponent $\alpha \simeq 2.56$ (see Table~\ref{tab:manna.alpha_results}).
      }
    \label{fig:manna.multilogcollapse3}
  \end{center}
\end{figure}

To demonstrate the universality of $\alpha$,
we consider four different PLG models in $d=2$
with a variety of initial conditions,
the details of which are given in Table~\ref{tab:manna.simdetails}.
The frequency distribution $N(r)$ 
for these four PLG models is shown in Figure~\ref{fig:manna.multilogcollapse3}.
The main plot shows the distributions recentered using
their respective means.  
The recentered distributions are indistinguishable to the eye
clearly indicating that the specifics of the growth
mechanisms are irrelevant and that 
the exponent $\alpha$ is a universal one.

The scaling theory developed in the previous section
relates the structure of a packing (its fractal dimension)
to the change of basic elements (critical radius, surface area, volume, etc.)
as a function of iteration.  In order to test the scaling
theory, as well as the specific prediction $\hat{\alpha}=d+(d+1)/(d+2)$
in higher dimensions, we now examine iterative
and structural scalings of the RAP model in $d=2$, 3, and 4.

We estimate $\alpha$ using a variety of means and the
results are summarized in Table~\ref{tab:manna.alpha_results}.
The agreement between the predicted value of $\alpha$ calculated using
$\Phi(n)$ and that calculated directly from the geometry
of the packing provides further proof that the scaling theory outlined
in the previous section is valid, regardless 
of the specific value of $\alpha$.
The current theory appears to improve with increasing
dimension, whereas the approximate theory of ABK only
holds in $d=2$.

\begin{table}[htbp!]
 \begin{center}
   \begin{tabular*}{\columnwidth}{l@{\extracolsep{\fill}}ccc@{\quad}}
     \toprule
     & $d=2$ & $d=3$ & $d=4$ \\
     \hline  
     $\Phi(n)$ & 2.564(1) & 3.733(2) & 4.833(2)\\
     $\Phi(r)$ & 2.58 & 3.73 & 4.83\\
     $S(r)$ &  2.55 & 3.74 & 4.86\\ 
     $N(r)$ &  2.53 & 3.70 & 4.79\\ 
     $\hat{\alpha} = d+[(d+1)/(d+2)]$ & 2.75 & 3.8 & 4.83333\\
     ABK theory & 2.553740 & 3.94505 & 4.99904\\
     \botrule
   \end{tabular*}
   \caption{
     Estimates of $\alpha$ using various numerical
     methods and comparisons to theory.
     All numerical results are from simulations containing $5\times 10^6$
     spheres.  Parentheses
     indicate $95\%$ confidence intervals on last digit.
     }
   \label{tab:manna.alpha_results}
 \end{center}
\end{table}

\begin{figure}[tbhp!]
  \begin{center}
    \epsfig{file=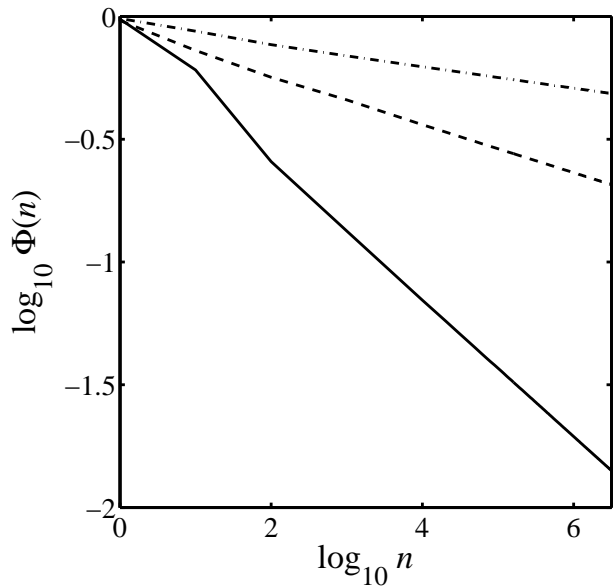,width=0.95\columnwidth}
    \caption{
      The decay of pore space volume 
      $\Phi(n)$ in $d=2$, 3 and 4 correspond to
      the solid, dashed, and dot-dashed lines respectively.  
      The fits to the power law decays are summarized in 
      Table~\ref{tab:manna.beta_results}.
      }
    \label{fig:manna.beta_numerics}
  \end{center}
\end{figure}

Both the current theory and that of ABK predict the
scaling of pore space as a function of iteration or time.
Evaluating the pore space decay $\Phi(n)$ in $d=2$ 
suggests that the agreement 
with ABK is most likely coincidental.  As evidenced
by the data in Fig.~\ref{fig:manna.beta_numerics} we
are able to establish $95\%$ confidence intervals that
exclude the predictions of ABK in $d=2$.  This is not
altogether surprising as ABK is an approximate theory
describing the collision of moving spheres whereas
in the RAP procedure all collisions are between 
a sphere and the presently jammed state.  
In higher dimensions, the theory of ABK fails, and 
the current theory becomes increasingly appropriate,
falling within the numerically defined
confidence intervals for $\beta$ in $d=4$.  
The results
are summarized in Table~\ref{tab:manna.beta_results}.

\begin{table}[btp!]
 \begin{center}
   \begin{tabular*}{\columnwidth}{l@{\extracolsep{\fill}}ccc@{\quad}}
     \toprule
     & $d=2$ & $d=3$ & $d=4$ \\
     \hline
     $\beta$ & 0.278(1) & 0.0975(2) & 0.0434(2)\\
     current theory \req{eq:manna.Phi(n)scaling} & 0.1429 & 0.07143 & 0.04348 \\
     ABK theory \req{eq:manna.abktheory} & 0.2872 & 0.01832 & $2.404\times{10^{-4}}$ \\
     \botrule
   \end{tabular*}
   \caption{
     The predicted exponent for the decline of
     volume fraction, $\Phi(n)\propto n^{-\beta}$. 
     The numerical estimates of $\beta$ are taken
     from simulations containing $5\times 10^6$ spheres.
     The decay of pore space as a function of iteration
     can be seen in Figure~\ref{fig:manna.beta_numerics}.
     Parenthesis indicate error on estimate of last digit.
     }
   \label{tab:manna.beta_results}
 \end{center}
\end{table}

\section{Conclusion}
\label{manna.conclusion}

The iterative nature of the random Apollonian packing 
model and its simple prescription
for packing spheres suggests that at sufficiently small scales,
the structure of packing-limited growth models is essentially
unaffected by initial conditions or dynamics.  
The scaling of pore space,
surface area, number, and critical radius $r_c$ are all interrelated
and can be expressed in terms of simple power laws using
only the dimension $d$ and the exponent $\alpha$.
Extensive numeric simulations demonstrate the validity of
the predicted upper bound $\hat{\alpha}$.  
Although $\hat{\alpha}$ appears to be an overestimate
of the actual value of $\alpha$ in $d=2$ and $d=3$, it
agrees with simulations in $d=4$, and presumably beyond.
A refined approximation of the insertion probability
$\Pins$ seems to be the best
approach by which to improve the estimate of $\alpha$.
The applicability of predictions of the present model
and its relevance to physical and biological problems
may lie in the process of balancing the 
idea of packing-limited growth
with other dynamical possibilities such
as aggregation, competition, and death.

\begin{acknowledgments}
This work was supported in part by NSF Grant No.\ EAR-9706220.
\end{acknowledgments}

\appendix

\end{document}